\documentclass[twocolumn,preprintnumbers,amsmath,amssymb,prl,floatfix]{revtex4}

\usepackage{graphicx}
\usepackage{dcolumn}
\usepackage{bm}

\newcommand{\eqb}{\begin{eqnarray}}
\newcommand{\eqe}{\end{eqnarray}}
\newcommand{\diff}{\textrm{d}}
\newcommand{\Eclassical}{E_{\rm cl}}
\newcommand{\Ecrit}{E_{\rm crit}}
\newcommand{\alphafine}{\alpha_{\rm f}}
\newcommand{\ehatvec}{\hat{{\bm e}}}
\newcommand{\betavec}{\bm{\beta}}
\newcommand{\lambdamicron}{\lambda_{\mu}}
\newcommand{\strength}{\hbox{$a$}}
\newcommand{\wsqcm}{\textrm{W\,cm}^{-2}}
\newcommand{\melec}{m_{\rm e}}
\begin{document} 

\title{Prolific pair production with high-power lasers}

\author{A.~R.~Bell}
\affiliation{Clarendon Laboratory, University of Oxford, 
Parks Road, Oxford OX1 3PU, UK} 
\affiliation{STFC Central Laser Facility, RAL, Didcot, OX11 0QX, UK}
\author{John~G.~Kirk}
\affiliation{Max-Planck-Institut f\"ur Kernphysik, Saupfercheckweg, 1, 
D-69117, Heidelberg, Germany}
\date{\today} 

\begin{abstract}
  Prolific electron-positron pair production is possible at laser intensities
  approaching 10$^{24}\,$Wcm$^{-2}$ at a wavelength of 1$\mu
  $m. An analysis of electron trajectories and interactions at the
  nodes ($B=0$) of two counter-propagating, circularly polarised laser
  beams shows that a cascade of $\gamma$-rays and pairs develops. The geometry
  is generalised qualitatively to linear polarisation and
  laser beams incident on a solid target.
\end{abstract}

\maketitle

High-power laser facilities have made dramatic progress recently, and 
the next few years may bring intensities of 
$10^{23}$--$10^{24}\,\wsqcm$ within reach. 
This naturally opens up new physics regimes \cite{kogaetal06,muelleretal08}.
The relativistic Lorentz
factor of an electron oscillating in vacuum in the electromagnetic field 
of 
a planar linearly polarised 
laser beam 
is 
$860\left(I_{24}\lambda _{\mu {\rm m}}^2\right)^{1/2}$
where $I_{24}$ is the laser 
intensity in $10^{24}\,\wsqcm$ and $\lambda_{\mu\rm m}$ is the laser
wavelength in micron. The corresponding peak
electric 
and magnetic fields are 
$2.7\times 10^{15}I_{24}^{1/2}\,$Vm$^{-1}$ 
and 
$91 I_{24}^{1/2}\,$GG. The  
Schwinger field $\Ecrit=1.3\times 10^{18}\,$Vm$^{-1}$
required for spontaneous electron-positron pair creation out of the vacuum 
would be attained
at a laser intensity of 
$2.3\times 10^{29}\,\wsqcm$ 
\cite{schwinger51,salaminetal06}.
Although this interesting regime 
is still far beyond projected laser intensities, 
several 
other strong-field
QED effects will soon be accessible to experiment.
In this {\em Letter\/} we show
how copious pair production by accelerated electrons interacting 
with the laser field can be achieved 
using laser
intensities $\sim10^{24}\,\wsqcm$. 
The key is to exploit the large 
transverse electromagnetic field seen by 
an electron when it experiences laser beams 
that are not propagating in parallel. We illustrate this effect
by computing the case of counter-propagating, circularly polarized beams.
The advantage offered by this configuration is analogous to the dramatic
increase in centre-of-mass energy when using colliding particle beams
instead of stationary targets. We argue that this advantage 
remains in less specific configurations such as 
tight focus and reflection from a solid surface.
Consequently, it may be possible to convert a large 
fraction of the laser energy into electron-positron pairs at a laser
intensity of $\sim10^{24}\,\wsqcm$ at approximately solid plasma 
density.

Relativistic electrons with Lorentz factor $\gamma$ 
moving perpendicular to a homogeneous 
magnetic field $B$ produce pairs if $\gamma B/B_{\rm crit}$
is greater than or of the order of unity, where,
$B_{\rm crit}=4.414\times10^4\,$GG is the magnetic equivalent of the 
Schwinger field $E_{\rm crit}$. The 
cross-sections for this and other relevant processes are well-known
\cite{erber66} and of interest also in astrophysics \cite{hardingli06}. 
Provided the electron trajectory can be approximated classically, 
these rates, when computed in a frame in which $\bm{E}\times\bm{B}=0$,  
are functions of the electric and magnetic fields only in the 
combination $E^2+B^2$ \cite{daughertylerche76,urrutia78}. 
Therefore, in a homogeneous electric field $\ehatvec E$, 
pair production occurs if the parameter
\eqb
\eta&=&\frac{\gamma E \sin \theta}{\Ecrit}
\label{etadef} 
\eqe  
is of order unity or larger, where 
$\theta$ is the angle between the electric
field and the electron momentum.
Pair
production by the trident process in 
which the electric field is provided by 
a high-Z nucleus and the Lorentz factor 
by accelerating electrons in the laser field,
has already been observed, but  
the process is relatively inefficient, and 
the yield achieved was $10^{-4}$ positrons for each fast
electron 
\cite{liangetal98,cowanetal99,nakashimaetal02}.  
Electron-positron
pairs have also been produced
by colliding $46.6\,$GeV electrons 
from a linear accelerator 
with an opposing laser beam, but this
produced a relatively modest number of pairs \cite{burkeetal97}.  

In a strong electromagnetic wave in vacuum, 
the Lorentz factor of an electron oscillates
about a value roughly equal to the strength parameter of the wave
\eqb
\strength&=&\frac{e E\lambda}{2\pi m c^2}
\nonumber
\\
&=&8.4\times 10^2 \left(I_{24}\lambda_{\mu{\rm m}}^2\right)^{1/2}
\label{strengthdef}
\eqe
\cite{lauetal03},
so that $\eta$ in Eq.~(\ref{etadef}) is approximately
$1.7 I_{24}\lambda_{\mu{\rm m}}$. This looks promising at first sight.
However, Eq.~(\ref{etadef}) assumes 
that $eE\sin\theta$ is the component of the particle's acceleration
perpendicular to its velocity, 
which is not the case in a laser field. In reality, 
an electron that is picked up by a single laser beam at initially low energy 
in the laboratory is accelerated on a trajectory that severely reduces the effective
value of $\eta$ below that in Eq.~(\ref{etadef}),
because the electric force is almost precisely cancelled
by that exerted by the magnetic field.

This can be understood in a way that brings
out the analogy with particle accelerators. In a 
plane electromagnetic wave in vacuum a charged particle has a
periodic trajectory in one special frame of reference 
(ignoring for
the moment radiation reaction) \cite{lauetal03}. 
This frame can be called the zero
momentum frame (ZMF).  
In it, all of 
the particle's phase-space variables are 
strictly periodic at the period of the wave, independently 
of its polarisation and waveform, and the particle energy 
oscillates around a value $\gamma mc^2\approx\strength mc^2$. 
In this frame, the electric and magnetic forces do not, in general,
cancel, 
and the perpendicular component of the acceleration is well-approximated by
$eE'$,  where $E'$ is the field
strength of the wave measured in the ZMF. Thus, 
the
importance of strong-field QED effects, such as pair-creation is
indeed determined by the parameter $\eta$, as defined in Eq.~(\ref{etadef}),
but computed in the ZMF, i.e., $\eta\approx \strength E'/\Ecrit$.  
In the case of a single laser beam
hitting a particle at rest, the ZMF does not coincide with the lab.\
frame, because the particle recoils. In fact, the ZMF moves in the
direction of propagation of the laser beam with a velocity
corresponding to a Lorentz factor equal to $\strength$ \cite{lauetal03}.  
The laser
frequency in the ZMF is thus red-shifted compared to the lab.\ frame. Because
the strength parameter $\strength$ is a Lorentz invariant, the reduced
wave frequency in the ZMF implies a reduced amplitude of the wave
field: $E'\approx E/\strength$. Consequently the (Lorentz invariant)
criterion for the importance of strong field QED effects becomes
$\strength E'/\Ecrit\approx E/\Ecrit>1$.  
In other words,
by using a single laser beam, the advantage gained over pure vacuum
effects by the relativistic oscillation of the electron is lost.

This is a consequence of the dual roles of accelerator and target
that are played by the laser beam. 
If the electron, instead of being initially at rest, is initially moving 
with a Lorentz factor $\gamma _{\rm init}$ that is much larger
than the strength parameter of the laser beam, as
in the experiment of Burke et al.\ \cite{burkeetal97}, the ZMF 
moves with almost the speed of the electron,
and $\eta\approx\gamma_{\rm init} E \sin \theta /\Ecrit$.
The initial Lorentz factor of the 
electron contributes to the threshold condition, but the
Lorentz factor due to oscillation in the laser field does not. In this case
the laser plays only the role of the target. 

However, as
in the case of intersecting particle beams, the situation can be
rescued if counter-propagating laser beams are employed. Then the
ZMF coincides with the lab.\ frame, and the importance of strong field
QED is again determined by Eq.~(\ref{etadef}).
A similar benefit is gained with a laser beam in tight focus --- which can
be decomposed into obliquely propagating plane waves ---
or with a beam in 
which a standing wave is set up when the laser encounters a dense
plasma.
Experimentally,
some of the 
most promising cases involve laser-solid interactions,
but the analysis of these is complicated.  Instead, we consider the 
theoretically simple case of pair production at the nodes 
($B=0$) of two
counter-propagating circularly polarised laser beams of 
equal intensity.  The argument can then be qualitatively generalised
to laser-solid interactions.  
Strong-field QED effects  
at the nodes of 
counter-propagating waves have been considered previously 
\cite{brezinitzykson70,dipiazzaetal06},
but only for a vacuum in which the 
threshold condition relates to $E$ rather than $\gamma E$.

Classically, the  electron equation of motion, including radiation 
reaction according to the Landau \& Lifshitz
prescription
\cite{landaulifshitz75}, is
\eqb
\frac{\diff(\gamma \betavec)}{\diff t}&=&
-\frac{e}{{\melec c}}(\bm{E}+\betavec \times \bm{B})-
\nonumber\\
&&\frac{2 e^4\gamma^2}{3 m^3 c^5} 
\betavec \left|\bm{E}+\betavec \times \bm{B}\right|_\bot^2
\label{llequation}
\eqe 
The terms
that have been omitted here are of order $\gamma^{-2}$. 
The final term of eq.~(\ref{llequation}) represents the drag
and energy loss due to 
radiative emission, to which pair production is related, and is
proportional to the square of that component of the Lorentz force
$\bm{E}+\betavec \times \bm{B}$ perpendicular to $\betavec$.  In the case
of a planar uni-directional wave, the reduction in the electric field
in the ZMF is equivalent to the near cancellation of $\bm{E}$ with
$\bm{\beta} \times \bm{B}$ in the laboratory frame.

Two counter-propagating laser beams produce a standing wave with nodes
at which $B=0$ and the electric field rotates in direction 
with constant amplitude. 
By symmetry, an electron placed exactly at
the node does not move in the direction of the waves, but 
performs circular motion with the
centripetal force provided by the electric field. 
The equation of motion (\ref{llequation}) then 
simplifies to 
\eqb
\frac{\diff \left(\gamma\betavec\right)}{\diff t}&=&\frac{e E}{mc}\left\lbrace
-\ehatvec -\frac{2}{3} \betavec\gamma^2\frac{E}{\Eclassical}\left[
1-\left(\betavec\cdot\ehatvec\right)^2\right]\right\rbrace
\label{eqmotionsimple}
\eqe
where $\hat{e}$ is the unit vector in the direction of the electric field
and we have defined the characteristic value $\Eclassical$ 
of the electric field in
classical electrodynamics: $\Eclassical=m^2c^4/e^3=\Ecrit/\alphafine$
($\alphafine$ is the fine-structure constant).
As can be seen from Eq.~(\ref{eqmotionsimple}), 
the radiation reaction force becomes important 
when $\gamma^2 E/\Eclassical\sim 1$, i.e.,
$\eta\sim\left(\gamma\alphafine\right)^{-1}$,  
a situation that is reached 
at laser intensities $\sim10^{23}\lambdamicron^{-4/3}\,\wsqcm$.
The Landau \& Lifshitz prescription for the radiation reaction term 
is valid up to 
$\eta\approx 1/\alphafine$
\cite{dipiazza08}
but quantum effects already intervene at $\eta\sim 1$
\cite{erber66}.

\begin{figure}
\includegraphics[width=0.5\textwidth]{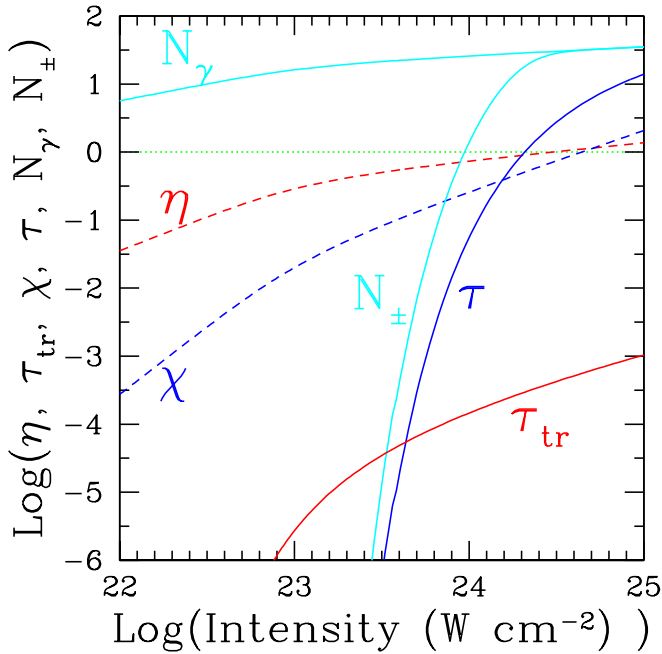}
\caption{%
\label{parameterspace}%
The parameters $\eta$ and $\chi$ --- see Eqs.~(\ref{etadef}) and
(\ref{chidef}) --- controlling the importance of electromagnetic
conversion by the accelerated electron and its curvature photon
respectively, as a function of laser intensity, for a laser wavelength
of $1\mu\textrm{m}$. Also plotted are the optical depth
$\tau$ of the curvature  
photon across one
wavelength, the number $N_\pm$
of pairs produced per electron in one laser
period by both this process and by that of
pair production by the trident process --- labelled $\tau_{\rm tr}$,
and the number $N_{\gamma}$
of curvature radiation photons produced per electron
per laser period.}
\end{figure}

Within a small fraction of a laser period, the electron trajectory
described by Eq.~(\ref{eqmotionsimple}) adjusts itself such that the
component of electric field parallel to $\betavec$ precisely
compensates the radiative losses, whilst the perpendicular component
enforces circular motion at the laser period, with Lorentz factor
$\gamma=\strength\sin \theta$. Combining these, and using the 
definition (\ref{etadef}) of the threshold parameter we can express
$E$ in terms of $\eta$.  
In an underdense plasma, the total electric field
$E$ is related to the intensity $I$ of each laser beam (separately)  
by $I=c E^2/(16\pi)$, which gives
\eqb
I_{24}&=&2.75\eta^4+0.28\left(\eta/\lambda_{\mu\rm m}\right)
\label{fluxeq}
\eqe
This relation is
plotted in Fig.~\ref{parameterspace} for a laser of wavelength
$1\,\mu $m.  In terms of these parameters, 
$\sin\theta=0.53\sqrt{\eta}\left(I_{24}\lambda_{\mu\rm m}\right)^{-1/2}$ --- 
at low intensity, the particle moves 
almost exactly perpendicularly to $\bm{E}$ 
and 
$\eta$ rises linearly with the laser intensity. However, when 
radiation reaction becomes important, this rise is slowed, and
$\eta=1$ is not achieved until $I_{24}=3$. The photons radiated 
because of the acceleration of the electron 
in the electric field of the laser --- which we term 
{\em curvature radiation}
\footnote{Frequently also called {\em bremsstrahlung}
\cite{erber66}, 
a term we reserve for radiation emitted in the electric field of a nucleus}
--- can
be described classically using the 
theory of synchrotron radiation. This predicts
that most radiated 
photons are emitted
with an energy 
\eqb
h\nu_{\rm s}&=&0.44\eta\gamma mc^2
\label{synchfreqdef}
\eqe where $\gamma mc^2=328\sqrt{\eta\lambda_{\mu{\rm m}}}\,$MeV is
the energy of the relativistic electron.  Because of quantum
effects analogous to the Klein-Nishina corrections to the Thomson
cross-section\cite{erber66,aharonian04}, 
the radiative energy loss does not proceed in the
continuous manner implied by Eq.~(\ref{eqmotionsimple}) when
$\eta>1$ and $I_{24}\gg1$. Neverthless, 
the classical trajectory is an adequate
approximation in the intensity range $10^{23}$--$10^{24}\wsqcm$, which
is of interest here, since the photon energy is significantly less
than the electron energy.

However, other important quantum effects are already present 
at intensities in this range. There are two 
processes that produce electron-positron pairs.  At low laser
intensities, the trident process dominates, 
in which an electron produces an
electron-positron pair via an intermediate virtual photon. 
In a homogeneous electric or magnetic field (a good
approximation when $\lambda_{\rm laser}\gg
h/mc=2.4\times10^{-6}\mu\textrm{m}$) the rate is given by
\cite{erber66,urrutia78}. Expressed as a production rate per electron
per laser period, it can be written $\tau_{\rm tr}=0.06\left
  (I_{24}\lambda_{\mu{\rm m}}^2\right
)^{1/2}\eta^{1/4}\textrm{exp}\left(-8/\sqrt{3\eta}\right)$ for
$\eta\ll1$, and, for $\eta>1$, it goes over to a slow logarithmic
increase. The precise form is plotted in Fig.~\ref{parameterspace}.

At higher intensities, the related process becomes important, 
in which the electron first
produces a real photon by curvature radiation, which subsequently
creates a pair.  However, to compare this to the
trident process one must specify the distance over which the real
photon is permitted to undergo conversion: if this is large, all
photons will convert into pairs, whereas if it is very short, none of
them will. The two processes are almost equal in rate if, at
$\eta\approx1$, the distance is chosen to be
$(\hbar/mc)(\Ecrit/E)=1.3\times10^{-4}I_{24}^{-1/2}\mu\textrm{m}$
\cite{erber66}.
However, in reality, this length is determined by the size of the region in 
which the laser beams overlap, which we 
conservatively assume to be $\lambda_{\rm laser}$.
This gives the process that involves a real photon as intermediary 
a substantial advantage. An additional, though less important, advantage arises 
because the propagation direction of 
the photon does not rotate, and, therefore, 
the perpendicular component of electric field it experiences is 
$\sim E$, rather than $ E \sin \theta$. 
The absorption coefficient is controlled by the parameter
\eqb
\chi&=&\frac{h\nu_{\rm s}E}{2mc^2\Ecrit} 
\label{chidef}
\eqe 
From Eq.~(\ref{synchfreqdef}) and writing $E$ in terms of $I_{24}$,
$\chi=0.42\eta^{3/2}\sqrt{I_{24}\lambda_{\mu{\rm m}}}$ and this 
function is plotted in Fig.~\ref{parameterspace}.  The 
photon optical depth to absorption in a path length $\lambda_{\rm laser}$ is
$\tau=12.8
\left(I_{24}\lambda_{\mu{\rm m}}^2\right)
\textrm{exp}\left[-4/\left(3\chi\right)\right]$, for
$\chi\ll1$, peaking at $\chi\approx8$ and falling off for larger
$\chi$ \cite{erber66}. It is also plotted in
Fig.~\ref{parameterspace}.
The total pair-production rate per electron
per laser period is the product of the photon absorption probability 
$1-\textrm{exp}\left(-\tau\right)$ in a length
$\lambda_{\rm laser}$ multiplied by the rate of production of photons by
curvature radiation. This quantity, together with the number of curvature
radiation photons 
emitted per
electron per laser period
(i.e., the energy radiated divided by $h\nu_{\rm s}$): 
$N_{\gamma}=6.42\alpha_{\rm f}\gamma$ is also 
shown in Fig.~\ref{parameterspace}.

Inspection of this figure shows that for laser intensities less than
roughly $I=3.3\times 10^{23}\,\wsqcm$, where $\eta=0.51$, pair
production is dominated by the trident process. At this intensity, each
electron in the zone where the laser beams overlap produces on average
$3\times 10^{-5}$ pairs in a single laser period. The curvature
radiation energy losses, which are $123\,$kW per electron, dominate
over pair production.  They are sufficient to damp the laser beams in
$1.8n_{23}^{-1}\,$fsec where $n_{23}$ is the electron density in
$10^{23}\,$cm$^{-3}$.  The total number of pairs produced 
in the absence of other energy losses is $7\times 10^4$
per Joule of laser energy.  

At
intensities above $I=3.3\times10^{23}\,\wsqcm$, the number of pairs
produced by photon-induced pair production rises steeply. 
These pairs are 
also accelerated and generate additional photons and pairs.  A cascade
should develop when $N_{\pm}\approx 1$, which occurs at 
$I_{24}\approx 1$ and $\eta\approx 0.7$.  
At this intensity, the laser power should be
divided roughly equally between photons and pairs with energy
$\sim 80\,$MeV per photon and per pair. This process is not sensitive to 
the number of electrons
initially in the interaction region. Complete conversion of laser energy
to photons and pairs implies the production of $\sim4\times 10^{10}$ pairs
per Joule of laser energy.  The precise conditions under which
a cascade is initiated are, however, sensitive to 
geometrical effects related to the intersection angle and the
intersection volume of the laser beams.

For simplicity of analysis we have assessed pair-production at the
nodes of counter-propagating circularly polarised waves and found that
the condition for pair production is 
$\strength E_\perp>\Ecrit$
where 
$\strength$ is the strength parameter, roughly equal to the 
Lorentz factor of an
electron oscillating in the electromagnetic wave.  This contrasts with
the much higher threshold in the case of an electron overtaken by a
planar uni-directional wave.  
The uni-directional plane wave is a special case, and the conditions
for prolific pair-production should be met at laser intensities $\sim
10 ^{24}\,\wsqcm$ away from the node or when the polarisation is not
circular, although the analysis is more complicated and numerical
factors of order unity will change the exact quantitative
result. Conditions for pair production may also occur at similar laser
intensities for a single laser beam incident on an overdense solid
target. The incident and reflected laser beams form
counter-propagating waves.  Even if the laser beam is substantially
absorbed rather than reflected, the electric field swells as the wave
passes into the plasma and the phase velocity differs from the speed
of light so that $\bm{E}$ and $\betavec\times\bm{B}$ 
are unlikely to cancel as in
the uni-directional case. Furthermore, if the laser beam is in tight
focus, as required for the highest intensities, it can be decomposed
into obliquely propagating plane waves. 
Another factor favouring pair production is
that the electrostatic field required for quasi-neutrality at a solid
surface stops the electrons moving freely in the direction of laser
propagation so that $\bm{E}$ and $\betavec\times\bm{B}$ 
once again cannot cancel, and the laser is  
able to play the 
role of particle accelerator and target simultaneously.

We predict that pair-production should be a standard feature of
laser-plasma interactions at intensities in the range $3 \times 10
^{23}-10 ^{24}$Wcm$^{-2}$ at a laser wavelength of 1$\mu $m. 
A significant number of pairs is produced at the lower end of this
intensity range and the number increases dramatically when the
intensity approaches $10 ^{24}\,\wsqcm$.  At intensities $\sim 10
^{24}\,\wsqcm$ a cascade sets in, producing an avalanche which
efficiently converts the laser energy into 
roughly equal numbers of pairs and $\gamma$-rays.
These predictions may be
tested using high-power lasers in the next few years.

\section*{Acknowledgments}
This research was carried out at 
the St John's College Research Centre, University of Oxford, and we
thank the members of the Centre for their warm hospitality. 


\begin{thebibliography}{18}
\expandafter\ifx\csname natexlab\endcsname\relax\def\natexlab#1{#1}\fi
\expandafter\ifx\csname bibnamefont\endcsname\relax
  \def\bibnamefont#1{#1}\fi
\expandafter\ifx\csname bibfnamefont\endcsname\relax
  \def\bibfnamefont#1{#1}\fi
\expandafter\ifx\csname citenamefont\endcsname\relax
  \def\citenamefont#1{#1}\fi
\expandafter\ifx\csname url\endcsname\relax
  \def\url#1{\texttt{#1}}\fi
\expandafter\ifx\csname urlprefix\endcsname\relax\def\urlprefix{URL }\fi
\providecommand{\bibinfo}[2]{#2}
\providecommand{\eprint}[2][]{\url{#2}}

\bibitem[{\citenamefont{{Koga} et~al.}(2006)\citenamefont{{Koga}, {Esirkepov},
  and {Bulanov}}}]{kogaetal06}
\bibinfo{author}{\bibfnamefont{J.}~\bibnamefont{{Koga}}},
  \bibinfo{author}{\bibfnamefont{T.~Z.} \bibnamefont{{Esirkepov}}},
  \bibnamefont{and} \bibinfo{author}{\bibfnamefont{S.~V.}
  \bibnamefont{{Bulanov}}}, \bibinfo{journal}{Journal of Plasma Physics}
  \textbf{\bibinfo{volume}{72}}, \bibinfo{pages}{1315} (\bibinfo{year}{2006}).

\bibitem[{\citenamefont{{M{\"u}ller} et~al.}(2008)\citenamefont{{M{\"u}ller},
  {di Piazza}, {Shahbaz}, {B{\"u}rvenich}, {Evers}, {Hatsagortsyan}, and
  {Keitel}}}]{muelleretal08}
\bibinfo{author}{\bibfnamefont{C.}~\bibnamefont{{M{\"u}ller}}},
  \bibinfo{author}{\bibfnamefont{A.}~\bibnamefont{{di Piazza}}},
  \bibinfo{author}{\bibfnamefont{A.}~\bibnamefont{{Shahbaz}}},
  \bibinfo{author}{\bibfnamefont{T.~J.} \bibnamefont{{B{\"u}rvenich}}},
  \bibinfo{author}{\bibfnamefont{J.}~\bibnamefont{{Evers}}},
  \bibinfo{author}{\bibfnamefont{K.~Z.} \bibnamefont{{Hatsagortsyan}}},
  \bibnamefont{and} \bibinfo{author}{\bibfnamefont{C.~H.}
  \bibnamefont{{Keitel}}}, \bibinfo{journal}{Laser Physics}
  \textbf{\bibinfo{volume}{18}}, \bibinfo{pages}{175} (\bibinfo{year}{2008}).

\bibitem[{\citenamefont{{Schwinger}}(1951)}]{schwinger51}
\bibinfo{author}{\bibfnamefont{J.}~\bibnamefont{{Schwinger}}},
  \bibinfo{journal}{Physical Review} \textbf{\bibinfo{volume}{82}},
  \bibinfo{pages}{664} (\bibinfo{year}{1951}).

\bibitem[{\citenamefont{{Salamin} et~al.}(2006)\citenamefont{{Salamin}, {Hu},
  {Hatsagortsyan}, and {Keitel}}}]{salaminetal06}
\bibinfo{author}{\bibfnamefont{Y.~I.} \bibnamefont{{Salamin}}},
  \bibinfo{author}{\bibfnamefont{S.~X.} \bibnamefont{{Hu}}},
  \bibinfo{author}{\bibfnamefont{K.~Z.} \bibnamefont{{Hatsagortsyan}}},
  \bibnamefont{and} \bibinfo{author}{\bibfnamefont{C.~H.}
  \bibnamefont{{Keitel}}}, \bibinfo{journal}{Physics Reports}
  \textbf{\bibinfo{volume}{427}}, \bibinfo{pages}{41} (\bibinfo{year}{2006}).

\bibitem[{\citenamefont{{Erber}}(1966)}]{erber66}
\bibinfo{author}{\bibfnamefont{T.}~\bibnamefont{{Erber}}},
  \bibinfo{journal}{Reviews of Modern Physics} \textbf{\bibinfo{volume}{38}},
  \bibinfo{pages}{626} (\bibinfo{year}{1966}).

\bibitem[{\citenamefont{{Harding} and {Lai}}(2006)}]{hardingli06}
\bibinfo{author}{\bibfnamefont{A.~K.} \bibnamefont{{Harding}}}
  \bibnamefont{and} \bibinfo{author}{\bibfnamefont{D.}~\bibnamefont{{Lai}}},
  \bibinfo{journal}{Reports of Progress in Physics}
  \textbf{\bibinfo{volume}{69}}, \bibinfo{pages}{2631} (\bibinfo{year}{2006}),
  \eprint{arXiv:astro-ph/0606674}.

\bibitem[{\citenamefont{{Daugherty} and {Lerche}}(1976)}]{daughertylerche76}
\bibinfo{author}{\bibfnamefont{J.~K.} \bibnamefont{{Daugherty}}}
  \bibnamefont{and} \bibinfo{author}{\bibfnamefont{I.}~\bibnamefont{{Lerche}}},
  \bibinfo{journal}{\prd} \textbf{\bibinfo{volume}{14}}, \bibinfo{pages}{340}
  (\bibinfo{year}{1976}).

\bibitem[{\citenamefont{{Urrutia}}(1978)}]{urrutia78}
\bibinfo{author}{\bibfnamefont{L.~F.} \bibnamefont{{Urrutia}}},
  \bibinfo{journal}{\prd} \textbf{\bibinfo{volume}{17}}, \bibinfo{pages}{1977}
  (\bibinfo{year}{1978}).

\bibitem[{\citenamefont{{Liang} et~al.}(1998)\citenamefont{{Liang}, {Wilks},
  and {Tabak}}}]{liangetal98}
\bibinfo{author}{\bibfnamefont{E.~P.} \bibnamefont{{Liang}}},
  \bibinfo{author}{\bibfnamefont{S.~C.} \bibnamefont{{Wilks}}},
  \bibnamefont{and} \bibinfo{author}{\bibfnamefont{M.}~\bibnamefont{{Tabak}}},
  \bibinfo{journal}{Physical Review Letters} \textbf{\bibinfo{volume}{81}},
  \bibinfo{pages}{4887} (\bibinfo{year}{1998}).

\bibitem[{\citenamefont{{Cowan} et~al.}(1999)\citenamefont{{Cowan}, {Perry},
  {Key}, {Ditmire}, {Hatchett}, {Henry}, {Moody}, {Moran}, {Pennington},
  {Phillips} et~al.}}]{cowanetal99}
\bibinfo{author}{\bibfnamefont{T.~E.} \bibnamefont{{Cowan}}},
  \bibinfo{author}{\bibfnamefont{M.~D.} \bibnamefont{{Perry}}},
  \bibinfo{author}{\bibfnamefont{M.~H.} \bibnamefont{{Key}}},
  \bibinfo{author}{\bibfnamefont{T.~R.} \bibnamefont{{Ditmire}}},
  \bibinfo{author}{\bibfnamefont{S.~P.} \bibnamefont{{Hatchett}}},
  \bibinfo{author}{\bibfnamefont{E.~A.} \bibnamefont{{Henry}}},
  \bibinfo{author}{\bibfnamefont{J.~D.} \bibnamefont{{Moody}}},
  \bibinfo{author}{\bibfnamefont{M.~J.} \bibnamefont{{Moran}}},
  \bibinfo{author}{\bibfnamefont{D.~M.} \bibnamefont{{Pennington}}},
  \bibinfo{author}{\bibfnamefont{T.~W.} \bibnamefont{{Phillips}}},
  \bibnamefont{et~al.}, \bibinfo{journal}{Laser Part.\ Beams}
  \textbf{\bibinfo{volume}{17}}, \bibinfo{pages}{773} (\bibinfo{year}{1999}).

\bibitem[{\citenamefont{{Nakashima} et~al.}(2002)\citenamefont{{Nakashima},
  {Cowan}, and {Takabe}}}]{nakashimaetal02}
\bibinfo{author}{\bibfnamefont{K.}~\bibnamefont{{Nakashima}}},
  \bibinfo{author}{\bibfnamefont{T.~E.} \bibnamefont{{Cowan}}},
  \bibnamefont{and} \bibinfo{author}{\bibfnamefont{H.}~\bibnamefont{{Takabe}}},
  in \emph{\bibinfo{booktitle}{Science of Superstrong Field Interactions}},
  edited by \bibinfo{editor}{\bibfnamefont{K.}~\bibnamefont{{Nakajima}}}
  \bibnamefont{and} \bibinfo{editor}{\bibfnamefont{M.}~\bibnamefont{{Deguchi}}}
  (\bibinfo{year}{2002}), vol. \bibinfo{volume}{634} of
  \emph{\bibinfo{series}{American Institute of Physics Conference Series}}, pp.
  \bibinfo{pages}{323--328}.

\bibitem[{\citenamefont{{Burke} et~al.}(1997)\citenamefont{{Burke}, {Field},
  {Horton-Smith}, {Spencer}, {Walz}, {Berridge}, {Bugg}, {Shmakov},
  {Weidemann}, {Bula} et~al.}}]{burkeetal97}
\bibinfo{author}{\bibfnamefont{D.~L.} \bibnamefont{{Burke}}},
  \bibinfo{author}{\bibfnamefont{R.~C.} \bibnamefont{{Field}}},
  \bibinfo{author}{\bibfnamefont{G.}~\bibnamefont{{Horton-Smith}}},
  \bibinfo{author}{\bibfnamefont{J.~E.} \bibnamefont{{Spencer}}},
  \bibinfo{author}{\bibfnamefont{D.}~\bibnamefont{{Walz}}},
  \bibinfo{author}{\bibfnamefont{S.~C.} \bibnamefont{{Berridge}}},
  \bibinfo{author}{\bibfnamefont{W.~M.} \bibnamefont{{Bugg}}},
  \bibinfo{author}{\bibfnamefont{K.}~\bibnamefont{{Shmakov}}},
  \bibinfo{author}{\bibfnamefont{A.~W.} \bibnamefont{{Weidemann}}},
  \bibinfo{author}{\bibfnamefont{C.}~\bibnamefont{{Bula}}},
  \bibnamefont{et~al.}, \bibinfo{journal}{Physical Review Letters}
  \textbf{\bibinfo{volume}{79}}, \bibinfo{pages}{1626} (\bibinfo{year}{1997}).

\bibitem[{\citenamefont{{Lau} et~al.}(2003)\citenamefont{{Lau}, {He},
  {Umstadter}, and {Kowalczyk}}}]{lauetal03}
\bibinfo{author}{\bibfnamefont{Y.~Y.} \bibnamefont{{Lau}}},
  \bibinfo{author}{\bibfnamefont{F.}~\bibnamefont{{He}}},
  \bibinfo{author}{\bibfnamefont{D.~P.} \bibnamefont{{Umstadter}}},
  \bibnamefont{and}
  \bibinfo{author}{\bibfnamefont{R.}~\bibnamefont{{Kowalczyk}}},
  \bibinfo{journal}{Physics of Plasmas} \textbf{\bibinfo{volume}{10}},
  \bibinfo{pages}{2155} (\bibinfo{year}{2003}).

\bibitem[{\citenamefont{{Brezin} and {Itzykson}}(1970)}]{brezinitzykson70}
\bibinfo{author}{\bibfnamefont{E.}~\bibnamefont{{Brezin}}} \bibnamefont{and}
  \bibinfo{author}{\bibfnamefont{C.}~\bibnamefont{{Itzykson}}},
  \bibinfo{journal}{\prd} \textbf{\bibinfo{volume}{2}}, \bibinfo{pages}{1191}
  (\bibinfo{year}{1970}).

\bibitem[{\citenamefont{{di Piazza} et~al.}(2006)\citenamefont{{di Piazza},
  {Hatsagortsyan}, and {Keitel}}}]{dipiazzaetal06}
\bibinfo{author}{\bibfnamefont{A.}~\bibnamefont{{di Piazza}}},
  \bibinfo{author}{\bibfnamefont{K.~Z.} \bibnamefont{{Hatsagortsyan}}},
  \bibnamefont{and} \bibinfo{author}{\bibfnamefont{C.~H.}
  \bibnamefont{{Keitel}}}, \bibinfo{journal}{Physical Review Letters}
  \textbf{\bibinfo{volume}{97}}, \bibinfo{pages}{083603}
  (\bibinfo{year}{2006}), \eprint{arXiv:hep-ph/0602039}.

\bibitem[{\citenamefont{{Landau} and {Lifshitz}}(1975)}]{landaulifshitz75}
\bibinfo{author}{\bibfnamefont{L.~D.} \bibnamefont{{Landau}}} \bibnamefont{and}
  \bibinfo{author}{\bibfnamefont{E.~M.} \bibnamefont{{Lifshitz}}},
  \emph{\bibinfo{title}{{The classical theory of fields}}}
  (\bibinfo{publisher}{Course of theoretical physics - Pergamon International
  Library of Science, Technology, Engineering and Social Studies, Oxford:
  Pergamon Press, 1975, 4th rev.engl.ed.}, \bibinfo{year}{1975}).

\bibitem[{\citenamefont{{di~Piazza}}(2008)}]{dipiazza08}
\bibinfo{author}{\bibfnamefont{A.}~\bibnamefont{{di~Piazza}}},
  \bibinfo{journal}{Lett.\ Math.\ Phys.} \textbf{\bibinfo{volume}{83}},
  \bibinfo{pages}{305} (\bibinfo{year}{2008}), \eprint{arXiv:0801.1751}.

\bibitem[{\citenamefont{{Aharonian}}(2004)}]{aharonian04}
\bibinfo{author}{\bibfnamefont{F.~A.} \bibnamefont{{Aharonian}}},
  \emph{\bibinfo{title}{{Very high energy cosmic gamma radiation : a crucial
  window on the extreme Universe}}} (\bibinfo{publisher}{Very high energy
  cosmic gamma radiation : a crucial window on the extreme Universe, by
  F.A.~Aharonian.~River Edge, NJ: World Scientific Publishing, 2004},
  \bibinfo{year}{2004}).

\end{thebibliography}
\end{document}